\documentclass[11pt,a4paper]{article}
\usepackage{amsmath,amsfonts,amssymb,amsthm,amstext,amscd,array}
\usepackage{tikz}
\usepackage{mathrsfs}
\usepackage{bbm}
\usepackage{bm}
\usepackage[arrow,matrix,curve]{xy}

\def\ii{{\,{\rm i}\,}}
\def\dd{{\rm d}}

\newcommand{\frg}{\mathfrak{g}}				

\newcommand{\eq}{\begin{equation}}
\newcommand{\eqend}{\end{equation}}
\newcommand{\eqa}{\begin{eqnarray}}
\newcommand{\nonueqa}{\begin{eqnarray*}}
\newcommand{\eqaend}{\end{eqnarray}}
\newcommand{\nonueqaend}{\end{eqnarray*}}

\newcommand{\bma}[1]{\begin{array}{#1}}
\newcommand{\ema}{\end{array}}
\newcommand{\bc}{\begin{center}}
\newcommand{\ec}{\end{center}}

\renewcommand{\thefootnote}{\fnsymbol{footnote}}
\newcommand{\newsection}{\setcounter{equation}{0}\section}

\newcommand{\complex}{{\mathbb C}} 
\newcommand{\real}{{\mathbb R}} 

\newif\ifold             \oldtrue

\def\e{{\,\rm e}\,}

\def\be{\begin{equation}}
\def\ee{\end{equation}}
\def\bea{\begin{eqnarray}}
\def\eea{\end{eqnarray}}
\def\bd{\begin{displaymath}}
\def\ed{\end{displaymath}}

\newcommand{\beq}{\begin{eqnarray}}
\newcommand{\eeq}{\end{eqnarray}}

\makeatletter
\newdimen\normalarrayskip              
\newdimen\minarrayskip                 
\normalarrayskip\baselineskip
\minarrayskip\jot
\newif\ifold             \oldtrue            
\def\arraymode{\ifold\relax\else\displaystyle\fi} 
\def\@arrayskip{\ifold\baselineskip\z@\lineskip\z@
     \else
     \baselineskip\minarrayskip\lineskip2\minarrayskip\fi}
\def\@arrayclassz{\ifcase \@lastchclass \@acolampacol \or
\@ampacol \or \or \or \@addamp \or
   \@acolampacol \or \@firstampfalse \@acol \fi
\edef\@preamble{\@preamble
  \ifcase \@chnum
     \hfil$\relax\arraymode\@sharp$\hfil
     \or $\relax\arraymode\@sharp$\hfil
     \or \hfil$\relax\arraymode\@sharp$\fi}}
\def\@array[#1]#2{\setbox\@arstrutbox=\hbox{\vrule
     height\arraystretch \ht\strutbox
     depth\arraystretch \dp\strutbox
     width\z@}\@mkpream{#2}\edef\@preamble{\halign \noexpand\@halignto
\bgroup \tabskip\z@ \@arstrut \@preamble \tabskip\z@ \cr}%
\let\@startpbox\@@startpbox \let\@endpbox\@@endpbox
  \if #1t\vtop \else \if#1b\vbox \else \vcenter \fi\fi
  \bgroup \let\par\relax
  \let\@sharp##\let\protect\relax
  \@arrayskip\@preamble}
\makeatother

\usepackage{hyperref}
\hypersetup{
unicode,pdftitle={},pdfauthor={}}

\begin{document}
\begin{titlepage}
\begin{flushright}

\baselineskip=12pt

HWM--14--01
\end{flushright}

\begin{center}

\vspace{2cm}

\baselineskip=24pt

{\Large\bf Topological gravity and transgression holography}

\baselineskip=14pt

\vspace{1cm}

{\bf Patricio Salgado}${}^1$\footnote{Email: \ {\tt pasalgad@udec.cl}}, {\bf Richard J. Szabo}${}^2$\footnote{Email: \ {\tt R.J.Szabo@hw.ac.uk}} and {\bf Omar Valdivia}${}^{1,2}$\footnote{Email: \ {\tt ov32@hw.ac.uk}}
\\[4mm]
\noindent ${}^1${\it Departamento de F\'isica\\ Universidad de Concepci\'on\\ Casilla 160-C,
  Concepci\'on, Chile}
\\[4mm]
\noindent ${}^2${\it Department of Mathematics\\ Heriot--Watt University\\ Colin Maclaurin Building,
  Riccarton, Edinburgh EH14 4AS, U.K.}\\ 
\noindent and {\it Maxwell Institute for
Mathematical Sciences, Edinburgh, U.K.}\\
\noindent and {\it The Tait Institute, Edinburgh, U.K.}
\\[30mm]

\end{center}

\begin{abstract}

\baselineskip=12pt

We show that Poincar\'e-invariant topological gravity in even
dimensions can be formulated as a transgression field theory in one higher dimension whose gauge connections are
associated to linear and nonlinear realizations of the Poincar\'e
group $ISO(d-1,1)$. The resulting theory is a gauged WZW model
whereby the transition functions relating gauge fields live in the
coset $\frac{ISO(d-1,1)}{SO(d-1,1)}$. The coordinate 
parametrizing the coset space is identified with the scalar field in
the adjoint representation of the gauge group of the even-dimensional
topological gravity theory. The supersymmetric extension 
leads to topological supergravity in two dimensions starting from a transgression field theory
which is
invariant under the supersymmetric extension of the Poincar\'e group
in three dimensions. We also apply this construction to a three-dimensional
Chern--Simons theory of gravity which is invariant under the Maxwell
algebra and obtain the corresponding WZW model.

\end{abstract}

\end{titlepage}
\setcounter{page}{2}

\newpage

{\baselineskip=12pt
\tableofcontents
}

\renewcommand{\thefootnote}{\arabic{footnote}}
\setcounter{footnote}{0}

\newpage

\newsection{Introduction and summary}

Gauge interactions, such as those which govern the Standard Model of
particle physics, are commonly based on a dynamical structure which
naturally encodes the property that nature should be invariant under a
group of transformations acting on each point of spacetime, i.e., a
local gauge symmetry. Although it is not guaranteed that gauge
invariant field theories are renormalizable, the only renormalizable
models describing nature are gauge theories, and thus the gauge symmetry principle seems to be a key ingredient in physically testable theories. 
The gravitational interaction, in contrast, has stubbornly
resisted quantization. General relativity seems to be the consistent
framework compatible with the idea that physics should be insensitive
to the choice of coordinates or the state of motion of any observer;
this is expressed mathematically as invariance under
reparametrizations or local diffeomorphisms. Although this invariance constitutes
a local symmetry, it does not qualify as a gauge symmetry. The reason
is that gauge transformations act on the fields while diffeomorphisms
act on their arguments, i.e., on the coordinates. A systematic way to
circumvent this obstruction is by using the tangent space
representation; in this framework gauge transformations constitute
changes of frames which leave the coordinates unchanged. However,
general relativity is not invariant under local
translations, except by a special accident in three spacetime
dimensions where the Einstein--Hilbert action is purely topological.
 
In refs.~\cite{Chamseddine:1989yz,Cha89,Cha90}
the classification of topological gauge theories for gravity is
presented. The natural gauge groups $G$ considered are the anti-de~Sitter
group $SO(d-1,2)$, the de~Sitter group $SO(d,1)$, and the Poincar\'e
group $ISO(d-1,1)$ in $d$ spacetime dimensions depending on the sign of the
cosmological constant: $-1,+1,0$ respectively. In odd dimensions
$d=2n+1$, the gravitational theories are constructed in terms of
secondary characteristic classes called
Chern--Simons forms. Chern--Simons forms are useful objects because they
lead to gauge invariant theories (modulo boundary terms). They also
have a rich mathematical structure similar to those of the (primary)
characteristic classes that arise in Yang--Mills theories: they are constructed in terms of a gauge potential which descends from a connection on a principal $G$-bundle.
 In even dimensions, there is no
natural candidate such as the Chern--Simons forms; hence in order to construct an invariant $2n$-form, the product of $n$ field strengths is not
sufficient and requires the insertion of a scalar multiplet $\phi^{a}$ transforming in the adjoint
representation of the gauge group $G.$ This requirement
ensures gauge invariance but it threatens the topological origin of
the theory. 

In this paper we show that even-dimensional topological gravity can be
formulated as a transgression field theory \cite{deAz95,Nak03} which
is invariant under the Poincar\'e group. The gauge connections are
considered as valued in both the Lie algebras associated to linear
and nonlinear realizations of the gauge group. The resulting theory is
a gauged Wess--Zumino--Witten (WZW) model
\cite{Wess:1967jq,Witten:1983tw} where the scalar field $\phi$ is now identified with the coset parameter of the nonlinear realization of the Poincar\'e group
$ISO(d-1,1)$. This identification allows the construction of topological
gravity as a holographic dual of a transgression gauge field theory in odd
dimensions. However, the transformation laws for the coset field break
translation invariance and therefore the residual symmetry is constrained
to the Lorentz subgroup $SO(d-1,1)$. This is not a huge obstruction in
the sense that one can restore the full Poincar\'e symmetry by
considering the coset fields as transforming in the adjoint
representation of the gauge group. By similar arguments we also
compute the transgression action for the $\mathcal{N}=1$ Poincar\'e
supergroup in three dimensions, and show that the resulting action is the one proposed in
\cite{Chamseddine:1989yz}; it would be
interesting to work out the generalizations of
this supergravity theory to
higher dimensions. Finally, we apply this construction to obtain a
gauged WZW model associated to the Maxwell algebra in
two dimensions; the resulting theory generalizes the topological
gravity theory proposed by \cite{Chamseddine:1989yz} and we show that,
in order to obtain the requisite invariant tensor associated to the Maxwell algebra, an $S$-expansion of the $\mathrm{AdS}$ algebra in three dimensions in required. There are various interesting extensions of our work that can be considered. For example, it would be interesting to construct a suitable transgression field theory in three dimensions for the Liouville theories considered in e.g.~\cite{Barnich:2013yka} which are invariant under the BMS algebra and which are dual to asymptotically flat Chern--Simons gravity. Moreover, it would interesting to construct the transgression field theories based on the AdS group $SO(2,2)=SL(2,\mathbb{R})\times SL(2,\mathbb{R})$ and more generally on gauge groups based on non-compact real forms of $SL(n,\mathbb{C})\times SL(n,\mathbb{C})$ in three dimensions which describe the couplings of higher spin $n$ fields to gravity~\cite{Campoleoni:2010zq}, and to study the dual WZW models; in such instances, however, our approach based on nonlinear realisations of the gauge group does not appear to be very tractable.

The structure of this paper is as follows. In Section \ref{sect2} we
review some aspects of topological gravity and its relation with
Lanczos--Lovelock theories of gravity. In Section \ref{sect3} we review
the formalism of nonlinear realizations of Lie groups and its
application to gravitational theories. Section \ref{sect4} introduces
transgression forms and presents the main results of this
investigation. As a representative example of how to incorporate
fermions into our construction, in
Section~\ref{sect5} we derive the supersymmetric extension of the
topological gravity action in the two-dimensional case. Section \ref{sect6} contains a brief application in which we construct the gauged WZW
model associated to the Maxwell algebra. Three appendices at the end of the paper contain
some technical details about the construction of Chern--Simons gravity
actions, our conventions for spinors, and the $S$-expansion method for Lie algebras.

\newsection{Topological gravity and Lanczos--Lovelock theory}\label{sect2}

\subsection{Topological gauge theories of gravity}

Topological gauge theories of gravity were classified in
refs. \cite{Chamseddine:1989yz,Cha89,Cha90}. The natural gauge groups
$G$ involved in the classification are given by
\begin{equation}
G \quad : \quad
\begin{tabular}
[c]{||l|l|l||}\hline\hline
\textrm{AdS} & $SO(d-1,2)$ & $\Lambda<0$\\\hline
\textrm{dS} & $SO(d,1)$ & $\Lambda>0$\\\hline
\textrm{Poincar\'e} & $ISO(d-1,1)$ & $\Lambda=0$\\\hline\hline
\end{tabular}
\end{equation}
depending on the spacetime dimension $d$ and the sign of the cosmological constant $\Lambda$. These
gauge groups are the smallest nontrivial choices which contain the
Lorentz symmetry $SO(d-1,1)$ as well as symmetries analogous to local
translations.

Throughout this paper, we shall let $\mathcal{P}$ denote a principal $G$-bundle
over a smooth manifold $\mathcal{M}$ of dimension $d$. Let $\mathcal{A}\in\Omega^{1}(
U,\frg)  $ be a local gauge potential with values in the Lie algebra
$\mathfrak{g}$ of $G$, obtained as the pull-back by a local section
$\sigma:U \rightarrow \mathcal{P}$, $U\subset\mathcal{M}$ of a one-form
connection $\theta\in\Omega^{1}(  \mathcal{P},\frg)  $ as
\begin{equation}
\mathcal{A}=\sigma^{\ast}\theta \ .
\end{equation}

In odd dimensions $d=2n+1$, the action for topological gravity is written in terms of a
Chern--Simons form defined by
\begin{equation}
\mathrm{S}^{\left(  2n+1\right)  }\left[  \mathcal{A}\right]  =\kappa \,
\int_{\mathcal{M}}\, \mathrm{L}_{\mathrm{CS}}^{\left(  2n+1\right)  }\left(
\mathcal{A}\right)  =\kappa\, \left(  n+1\right) \, \int_{\mathcal{M}}
\ \int_{0}%
^{1} \, \dd t \ \left\langle \mathcal{A} \wedge\left(  t\,
    \dd\mathcal{A}+t^{2} \, 
\mathcal{A\wedge A}\right)  ^{n}\right\rangle \ . \label{cspoinc}%
\end{equation}
Here $\kappa\in\real$ is a constant and $\left\langle - \right\rangle
:\mathfrak{g}^{\otimes(n+1)} \rightarrow\mathbb{R}
$ is a $G$-invariant symmetric polynomial of rank $n+1$ which
is determined once an explicit presentation for $\mathfrak{g}$ is
chosen. Note that the Chern--Simons form $\mathrm{L}_{\mathrm{CS}}^{\left(  2n+1\right)  }\left(
\mathcal{A}\right)$ is not globally-defined (unless $\mathcal{P}$ is
trivial) and the gauge theory
specified by eq.~(\ref{cspoinc}) is regarded as defined by a
sheaf of local Lagrangians.

In even dimensions there is no topological candidate such as the Chern--Simons form. In fact, the
exterior product of $n$ field strengths makes the required $2n$-form in a
$2n$-dimensional spacetime, but in order to obtain a gauge invariant
differential $2n$-form, a scalar multiplet $\phi^{a}$ with $a=1,\ldots
,2n+1$ transforming in the adjoint representation of the gauge group must be
added and the action is given by
\begin{equation}
\mathrm{S}^{\left(  2n\right)  }\left[  \mathcal{A},\phi\right]  =\kappa\,
\int_{\mathcal{M}}\, \big\langle \mathcal{F}^n \wedge
\phi\big\rangle \ . \label{maineq}%
\end{equation}
Here $\mathcal{F}=\dd\mathcal{A+A\wedge A}$ is the curvature two-form
associated to the gauge potential $\mathcal{A}$. Note that here the
Lagrangian $\big\langle \mathcal{F}^n \wedge
\phi\big\rangle$ is a global differential form on $\mathcal{M}$. This topological action has interesting
applications; for instance, in two dimensions it describes the Liouville theory
of gravity from a local Lagrangian \cite{D'Hoker:1983ef,D'Hoker:1983is}.

\subsection{Lanczos--Lovelock gravity}

The most general Lagrangian in $d$ dimensions which is compatible with the
Einstein--Hilbert action for gravity is a polynomial of degree $\left[  d/2\right]  $ in
the curvatures known as the Lanczos--Lovelock Lagrangian
\cite{Lan38,Lov71,Cno05,Des05,Mar91}. Lanczos--Lovelock theories share
the same fields, symmetries and local degrees of freedom of General
Relativity. The Lagrangian is built from the
vielbein $e^{a}$ and the spin connection $\omega^{ab}$ via the Riemann
curvature two-form $R^{ab}=\dd\omega^{ab}+\omega
_{~c}^{a}\wedge\omega^{cb}$, leading to the action
\begin{equation}
\mathrm{S}^{(d)}_{\mathrm{LL}} =\int_{\mathcal{M}} \ \sum\limits_{p=0}^{\left[
d/2\right]  }\, \alpha_{p}\, \epsilon_{a_{1}\cdots a_{d}}\,
R^{a_{1}a_{2}}\wedge \cdots \wedge
R^{a_{2p-1}a_{2p}}\wedge e^{a_{2p+1}}\wedge \cdots \wedge e^{a_d} \ .
\end{equation}
Here $\alpha_{p}$ are arbitrary parameters that cannot be fixed from first
principles. However, in ref.~\cite{Tro99} it is shown that by requiring the
equations of motion to uniquely determine the dynamics for as many
components of the independent fields as possible, one can fix $\alpha_{p}$ (in any
dimension) in terms of the gravitational and cosmological constants.

In $d=2n$ dimensions the parameters $\alpha_{p}$ are given by%
\begin{equation}
\alpha_{p}=\alpha_{0}\, \left(  2\gamma\right)  ^{p}\, \binom{n}{p} \label{coeffeven}
\end{equation}
and the Lagrangian takes a Born--Infeld form. The Lanczos--Lovelock action
constructed in this dimension is only invariant under the Lorentz symmetry
$SO(2n-1,1)$.
In odd dimensions $d=2n+1$ the coefficients are given by
\begin{equation}
\alpha_{p}=\alpha_{0}\, \frac{(2n-1)\, (2\gamma)  ^{p}%
}{2n-2p-1}\, \binom{n-1}{p} \ . \label{coeffodd}
\end{equation}
Here
\begin{equation}
\alpha_{0}=\frac{\kappa}{d\, l^{d-1}} \qquad \mbox{and} \qquad \gamma=-\mathrm{sgn}(
\mathrm{\Lambda}) \, \frac{l^{2}}{2}
\end{equation}
with $\kappa$ an arbitrary dimensionless constant, and $l$ is a length parameter related to the
cosmological constant by
\begin{equation}
\mathrm{\Lambda}=\pm\, \frac{(d-1)\, (d-2)}{2l^{2}} \ .
\end{equation}

With this choice of coefficients, the
Lanczos--Lovelock Lagrangian for $d=2n+1$
coincides exactly with a Chern--Simons form for the $\mathrm{AdS}$
group $SO(2n,2)$. This means that the exterior derivative of the Lanczos--Lovelock Lagrangian corresponds to a $2n+2$-dimensional Euler density. This is the reason
why there is no analogous construction in even dimensions: There are no known
topological invariants in odd dimensions which can be constructed
in terms of exterior products of curvatures alone. However, in ref.~\cite{Sal05} a Lanczos--Lovelock theory
genuinely invariant under the $\mathrm{AdS}$ group, in any dimension, is
proposed. The construction is based on the Stelle--West mechanism \cite{Ste80,Grignani:1991nj},
which is an application of the theory of nonlinear realizations of Lie groups
to gravity.

\newsection{Nonlinear realizations of Lie groups}
 \label{sect3}

\subsection{Nonlinear gauge theories}

Nonlinear realizations of Lie groups were introduced in refs.~\cite{Coleman:1969sm,Callan:1969sn}.
Following these references, let $G$ be a (super-)Lie group of
transformations of dimension $n$
and $\mathfrak{g}$ its Lie algebra. Let $H$ be a stability subgroup of $G$
of dimension $n-d$ whose Lie algebra $\mathfrak{h}$ is generated by
$\left\{  \mathsf{V}_{i}\right\}  _{i=1}^{n-d}$. Let us denote by
$\mathfrak{p}$ the vector subspace generated by the remaining generators of
$\mathfrak{g}$, denoted $\left\{  \mathsf{P}_{l}\right\}
_{l=1}^{d}$, such that there is a vector space decomposition $\mathfrak{g}=\mathfrak{h}\oplus\mathfrak{p}$. Since
$\mathfrak{h}$ is a subalgebra, one has $\left[  \mathfrak{h,h}\right]
\subset\mathfrak{h}$. We will further assume that $\mathfrak{p}$ can be
chosen in such a way that it defines a representation of $H$, so that
$\left[  \mathfrak{h},\mathfrak{p}\right]  \subset\mathfrak{p}$.
With this decomposition, any element $g_{0}\in G$ can always be uniquely
written as%
\begin{equation}
g_{0}=\e^{\zeta\cdot \mathsf{P}}\, h \label{nonl1}%
\end{equation}
where $h\in H$ and $\e^{\zeta\cdot \mathsf{P}}=\e^{\zeta
^{l}\, \mathsf{P}_{l}}\in G/H$ with $l=1,\ldots ,d$. The local coordinates $\zeta$
parametrize the coset space $G/H$. By virtue of eq.~$\left(  \ref{nonl1}%
\right)  $, the action of $g_{0}$ on the coset space $G/H$ is given by%
\begin{equation}
g_{0}\, \e^{\zeta\text{\textperiodcentered}\mathsf{P}}=\e^{\zeta^{\prime
}\text{\textperiodcentered}\mathsf{P}}\, h_{1} \ . \label{nonl2}%
\end{equation}
This expression allow us to obtain $\zeta^{{\prime}}$ and $h_{1}$ as certain
nonlinear functions of $g_{0}$ and $\zeta$,
\begin{equation}
\zeta^{{\prime}}=\zeta^{{\prime}}(  g_{0},\zeta)  \qquad \mbox{and}
\qquad h_{1}=h_{1}(  g_{0},\zeta)  \ . \label{nonl4}%
\end{equation}
For $g_{0}$ close to the identity, eq.~$\left(  \ref{nonl2}\right)  $ reads%
\begin{equation}
\e^{-\zeta\text{\textperiodcentered}\mathsf{P}}\, \left(
  g_{0}-1\right) \,
\e^{\zeta\text{\textperiodcentered}\mathsf{P}}-\e^{-\zeta
\text{\textperiodcentered}\mathsf{P}}\, \delta \e^{\zeta\text{\textperiodcentered
}\mathsf{P}}=h_{1}-1 \ , \label{nonl3}%
\end{equation}
allowing us to obtain the variation $\delta\zeta=\zeta'-\zeta$ under the infinitesimal action of $G$.
Note that if we restrict $G$ to the subgroup $H$, the nonlinear representation
becomes linear: If $g_{0}=h_{0}\in H$, then eq.~$\left(  \ref{nonl2}\right)  $ takes
the form%
\begin{equation}
\e^{\zeta^{\prime}\text{\textperiodcentered}\mathsf{P}}\,
h_{1}=h_{0}\, 
\e^{\zeta\text{\textperiodcentered}\mathsf{P}}=\big(  h_{0}\, \e^{\zeta
\text{\textperiodcentered}\mathsf{P}}\, h_{0}^{-1}\big)\,  h_{0} \label{nonl8}%
\end{equation}
and since $\left[  \mathfrak{h},\mathfrak{p}\right]  \subset\mathfrak{p}$ the
term $h_{0}\, \e^{\zeta\text{\textperiodcentered}\mathsf{P}}\,
h_{0}^{-1}$ is an exponential in the generators of $\mathfrak{p}$, implying
\begin{align}
h_{1}  &  =h_{0} \ , \label{nonl6}\\[4pt]
\e^{\zeta^{\prime}\text{\textperiodcentered}\mathsf{P}}  &  =h_{0}\,
\e^{\zeta\text{\textperiodcentered}\mathsf{P}}\, h_{0}^{-1} \ , \label{nonl7}%
\end{align}
where the transformation from $\zeta$ to $\zeta^{\prime}$ in eq.~$\left(
\ref{nonl7}\right)  $ is linear.
On the other hand, if
\begin{equation}
g_{0}=\e^{\zeta_{0}\text{\textperiodcentered}\mathsf{P}} \ \in \ G/H
\ , \label{nonl9}%
\end{equation}
then eq.~$\left(  \ref{nonl2}\right)  $ becomes
\begin{equation}
\e^{\zeta_{0}\text{\textperiodcentered}\mathsf{P}}\, \e^{\zeta
\text{\textperiodcentered}\mathsf{P}}=\e^{\zeta^{\prime}%
\text{\textperiodcentered}\mathsf{P}}\, h_{1} \label{nonl10}%
\end{equation}
which is a nonlinear transformation in the coset coordinate $\zeta$.

The construction of Lagrangians which are invariant under local gauge transformations
usually involves the introduction of a set of gauge fields associated with the
generators of $\mathfrak{g}$. Here, as in the case of linear representations,
a nonlinear gauge potential $\mathcal{\bar{A}}$ must be introduced in order to
guarantee that the derivatives of the fields $\zeta,\bar{\varphi}$ transform
covariantly, where $\bar\varphi$ are coordinates on the subgroup $H$. The linear gauge potential $\mathcal{A}$ can be naturally
decomposed into gauge fields associated to $H$ and $G/H$ as
\begin{equation}
\mathcal{A}=v^{i}\, \mathsf{V}_{i}+p^{l}\, \mathsf{P}_{l} \ .
\end{equation}
Introducing the nonlinear gauge potential $\mathcal{\bar{A}}$, we can
now write the nonlinear gauge fields
\begin{equation}
\mathcal{\bar{A}}=\bar{v}{}^{i}\, \mathsf{V}_{i}+\bar{p}{}^{l}\, \mathsf{P}_{l} \ .
\end{equation}
The
linear and nonlinear gauge potentials are related by \cite{Coleman:1969sm,Callan:1969sn}
\begin{equation}
\bar{v}{}^{i}\,\mathsf{V}_{i}+\bar{p}{}^{l}\, \mathsf{P}_{l}=\e^{\zeta\text{\textperiodcentered
}\mathsf{P}}\, \big(  \dd+v^{i}\, \mathsf{V}_{i}+p^{l}\, \mathsf{P}_{l}\big)  \e^{-\zeta
\text{\textperiodcentered}\mathsf{P}} \ . \label{gfield}
\end{equation}
This relation has exactly the form of a gauge
transformation by $\e^{-\zeta \cdot \mathsf{P}}\in G/H$. The transformation
relations for $\bar{v}=\bar{v}(  \zeta,\dd\zeta)  $, $\bar{p}=\bar{p}(  \zeta
,\dd\zeta)  $ are obtained using eqs.~$\left(  \ref{nonl1},\ref{nonl2}%
\right)  $ and are given by%
\begin{align}
\bar{v}{}^{\prime}  & =h_{1}^{-1}\, \bar{v}\, h_{1} \ , \label{nonl11}\\[4pt]
\bar{p}{}^{\prime}  & =h_{1}^{-1}\, \bar{p} \, h_{1}+h_{1}^{-1}\, \dd h_{1}\label{nonl12}%
\end{align}
with $h_{1}(  \zeta,\zeta_{0})  \in H$. Eqs.~$\left(  \ref{nonl11}%
,\ref{nonl12}\right)  $ show that the nonlinear fields $\bar{v}=\bar{v}(
\zeta,\dd\zeta)  $ and $\bar{p}=\bar{p}(  \zeta,\dd\zeta)  $ transform as
a tensor and as a connection respectively under the action of $h_{1}(
\zeta,\zeta_{0})  \in H$. Since $h_{1}$ depends on $\zeta$, any $H$-invariant expression written in terms of $v$ and $p$ will be automatically
invariant under the full group $G$, provided one replaces the linear gauge
fields $v$ and $p$ by their
nonlinear versions $\bar{v}$ and $\bar{p}$. We now make use of the nonlinear
gauge fields
and their properties to define the covariant derivative respect to the group
$G$ as
\begin{equation}
D_{\bar{p}} := \dd+\bar{p} \ ,
\end{equation}
and the corresponding curvature two-form whose components are given by%
\begin{equation}
T =D_{\bar{p}}\bar{v}\qquad \mbox{and} \qquad 
R =\dd\bar{p}+\bar{p}\wedge\bar{p} \ .
\end{equation}

\subsection{SWGN formalism}

The Stelle--West--Grignani--Nardelli (SWGN) formalism \cite{Ste80,Grignani:1991nj}
is an application of the theory of nonlinear realizations of Lie groups to
gravity. In particular, it allows the construction of the
Lanczos--Lovelock theory of gravity which is genuinely
invariant under the anti-de~Sitter group $G=SO(d-1,2)$. This model is discussed in ref.~\cite{Sal03b} and it is described by the action
\begin{equation}
\mathrm{S}_{\mathrm{SW}}^{\left(  d\right)  }=\int_{\mathcal{M}} \
\sum\limits_{p=0}^{\left[  d/2\right]  } \,
\alpha_{p}\, \epsilon_{a_{1} \cdots a_{d}}\, \bar{R}{}^{a_{1}a_{2}}\wedge
\cdots \wedge\bar{R}{}^{a_{2p-1}a_{2p}}\wedge\bar{e}{}^{a_{2p+1}}%
\wedge \cdots \wedge\bar{e}{}^{a_{d}} \ .\label{llact}%
\end{equation}
Here $\bar{R}{}^{ab}$ and $\bar{e}{}^{a}$ are nonlinear gauge
fields and the coefficients $\alpha_{p}$ are given by either eq.~$(\ref{coeffeven})$ or eq.~$(\ref{coeffodd})$ depending on the dimension of the spacetime. Using eq.~$(\ref{gfield})$ we get
\begin{align}
\bar{e}{}^{a} & =\Omega_{~b}^{a}( \cosh z) \, e^{b}+\Omega
_{~b}^{a}\Big(\, \frac{\sinh z}{z}\, \Big) \, D_{\omega}\phi^{b} \ , \label{xor1}\\[4pt]
\bar{R}{}^{ab} & =\dd\bar{\omega}{}^{ab}+\bar{\omega}{}_{~c}^{a}\wedge
\bar{\omega}{}^{cb} \ , \label{xor2}\\[4pt]
\bar{\omega}{}^{ab} & =\omega^{ab}+\frac{\sigma}{l^{2}}\, \Big(\,
\frac{\sinh z}{z}\, \big(  \phi^{a}\, e^{b}-\phi^{b}\, e^{a}\big)
+\frac{\cosh z-1}{z^{2}} \, \big( \phi^{a}\,
D_{\omega}\phi^{b}-\phi^{b}\, D_{\omega}\phi
^{a}\big) \, \Big) \ , \label{xor3}%
\end{align}
where $e^a$ and $\omega^{ab}$ are the usual vielbein and spin connection, respectively. We
have defined
\begin{align}
D_{\omega}\phi^a & := \dd\phi^a + \omega_{~b}^{a} \, \phi^b \ ,
\nonumber \\[4pt]
z  & := \frac{\phi}{l}=\frac{\sqrt{\phi^{a}\, \phi_{a}}}{l} \ ,
\nonumber \\[4pt]
\Omega_{~b}^{a}(  u)    & := u\, \delta_{~b}^{a}+\left(
1-u\right) \, \frac{\phi^{a}\, \phi_{b}}{\phi^{2}} \ ,
\end{align}
where $l$ is the radius of curvature of AdS and $\phi^{a}$ are the \textrm{AdS} coordinates
 which parametrize the coset space $
\frac{SO(d-1,2)}{SO(d-1,1)} $. In this scheme, this coordinate carries no
dynamics as any value that we pick for it is equivalent to a gauge
fixing condition
which breaks the symmetry from \textrm{AdS} to the Lorentz subgroup. This is
best seen using the equations of motion; they are the same as those
for the ordinary Lanczos--Lovelock theory where the vielbein $e^{a}$ and the spin
connection $\omega^{ab}$ are replaced by their nonlinear versions
$\bar{e}{}^{a}$ and $\bar{\omega}{}^{ab}$ given in eqs.~$\left(  \ref{xor1}%
,\ref{xor3}\right)  $. 

In odd
dimensions $d=2n+1$, the Chern--Simons action written in terms of the linear
gauge fields $e^a$ and $\omega^{ab}$ with values in the Lie algebra of
$SO(2n,2)$ differs only by a boundary term from that written using
the nonlinear gauge fields $\bar{e}{}^a$
and $\bar{\omega}{}^{ab}$. This is by virtue of eq.~$(\ref{gfield})$ which has the form of a gauge transformation
\begin{equation}
\mathcal{A} \ \longmapsto \ \mathcal{\bar{A}}=g^{-1}\, \left(  \dd+\mathcal{A}%
\right)  \, g \label{gtc}
\end{equation}
with $g=\e^{-\phi^{a}\, \mathsf{P}_{a}} \in  \frac{SO(2n,2)}%
{SO(2n,1)}  $. Alternatively, since $\mathcal{\bar{F}}=g^{-1}\,
\mathcal{F}\, g$ we have
\begin{equation}
\dd\, \mathrm{L}_{\mathrm{CS}}^{\left(  2n+1\right)  }\left(
  \mathcal{\bar{A}
} \, \right)  =\left\langle \mathcal{\bar{F}}^{n+1}\right\rangle =\left\langle
\mathcal{F}^{n+1}\right\rangle =\dd\, \mathrm{L}_{\mathrm{CS}}^{\left(
2n+1\right)  }\left(  \mathcal{A}\right)
\end{equation}
and hence both Lagrangians may locally differ only by a total
derivative.

\subsection{Poincar\'e gravity}

In odd spacetime dimensions $d=2n+1$, Poincar\'e gravity is a
Chern--Simons theory for the gauge group $G=ISO(2n,1)$. This group can
be obtained by performing an In\"{o}n\"{u}--Wigner contraction of
the \textrm{AdS} group in odd dimensions $SO(2n,2)$. The bulk part
of the Lagrangian can still be recovered in the limit $l\to\infty$ from the Lanczos--Lovelock
series in the case $d=2n+1$ and $p=n$. However, there is an extra
boundary term which arises once the computation of the relevant Chern--Simons
form is carried out; see Appendix \ref{app1} for
details and conventions. The resulting Lagrangian is then given by
\begin{align}
\mathrm{L}_{\mathrm{CS}}^{\left(  2n+1\right)  }\left(  \mathcal{A}\right)
=&\ \epsilon_{a_{1}\cdots a_{2n+1}}\, R^{a_{1}a_{2}}\wedge \cdots \wedge R^{a_{2n-1}a_{2n}}\wedge
e^{a_{2n+1}}\nonumber\\
& -n\, \dd\int_{0}^{1}\, \dd t \ t^{n}\, \epsilon_{a_{1}
  \cdots a_{2n+1}}\, R_{t}^{a_{1}a_{2}}%
\wedge \cdots \wedge R_{t}^{a_{2n-3}a_{2n-2}}\wedge\omega^{a_{2n-1}a_{2n}}\wedge
e^{a_{2n+1}}\label{cslin}%
\end{align}
where $R_{t}^{ab}=\dd\omega^{ab}+t\, \omega_{~c}^{a}\wedge\omega^{cb}$.
Under infinitesimal local gauge transformations with parameter
$\lambda=\frac{1}{2}\, \kappa^{ab}\, \mathsf{J}_{ab}+\rho^a \, \mathsf{P}_a$, the gauge fields transform as 
\begin{equation}
\delta e^{a}    =-D_{\omega}\rho^{a}+\kappa_{~b}^{a}\, e^{b} \qquad
\mbox{and} \qquad 
\delta\omega^{ab}    =-D_{\omega}\kappa^{ab} \ ,
\end{equation}
and these transformations leave eq.~($\ref{cslin}$) invariant modulo
a total derivative.

The nonlinear Lagrangian can be obtained using eqs.~$(\ref{xor1},
\ref{xor3})$ in the limit $l\rightarrow\infty$ and substituting into eq.~$(\ref{cslin})$ to obtain
\begin{align}
\mathrm{L}_{\mathrm{CS}}^{\left(  2n+1\right)  }\left(  \mathcal{\bar{A}%
}\, \right) = & \ \epsilon_{a_{1} \cdots a_{2n+1}}\, R^{a_{1}a_{2}}\wedge \cdots \wedge R^{a_{2n-1}%
a_{2n}}\wedge\left(
e^{a_{2n+1}}+D_{\omega}\phi^{a_{2n+1}}\right) \nonumber \\
& -n\, \dd\int_{0}^{1}\, \dd t \ t^{n}\,
\epsilon_{a_{1}\cdots a_{2n+1}}\, R_{t}^{a_{1}a_{2}}%
\wedge \cdots \wedge R_{t}^{a_{2n-3}a_{2n-2}}\wedge\omega^{a_{2n-1}a_{2n}}
\nonumber \\ & \hspace{5cm} \wedge\left(
e^{a_{2n+1}}+D_{\omega}\phi^{a_{2n+1}}\right) \ . \label{csnonlin} 
\end{align}
The gauge transformations for the coset field $\phi$ can be obtained
from eq.~$(\ref{nonl3})$ using $g_{0}-1=-\phi^a \, \mathsf{P}_a$. In
this case one shows that under local Poincar\'e translations the coset
field $\phi$ transforms as $\delta \phi^a=\rho^a$. One can
directly check, as in the case of the linear Lagrangian, that
eq.~$(\ref{csnonlin})$ remains unchanged under gauge transformations
up to a total derivative.

\newsection{Topological gravity as a transgression field theory} \label{sect4}

\subsection{Transgression forms as global Lagrangians}

In this section we show that the topological action for gravity in
$2n$ dimensions given in eq.~$\left(  \ref{maineq}\right)  $ can be obtained from a
$\left(  2n+1\right)$-dimensional transgression field theory which is genuinely invariant under the
Poincar\'e group $G= ISO(2n,1)$. 
 Transgression forms are generalizations of Chern--Simons
forms. They are gauge invariant objects and use, in addition to the
gauge potential $\mathcal{A}$, a
second Lie algebra valued gauge potential $\mathcal{\bar{A}}$. Due to their full
invariance property they are good candidates for the construction of action
principles by regarding $\mathcal{A}$ and $\mathcal{\bar{A}}$ as fundamental fields \cite{Bor03,Bor05,Mor06a,Mor05}.

Let $\mathcal{A}$ and $\mathcal{\bar{A}}$ be two $\mathfrak{g}$-valued
connections. The transgression field theory is defined by the action
\begin{equation}
\mathrm{S}_{\mathrm{T}}^{\left(  2n+1\right)  }\left[  \mathcal{A}%
,\mathcal{\bar{A}}\, \right]  =\kappa\, \int_{\mathcal{M}}\, Q_{\mathcal{A}%
\leftarrow\mathcal{\bar{A}}}^{\left(  2n+1\right)  }=\kappa\, \left(
n+1\right)\, 
\int_{\mathcal{M}} \ \int_{0}^{1}\, \dd t \ \big\langle
\big(\mathcal{A}-\bar{\mathcal{A}}\, \big) \wedge
\mathcal{F}_{t}^{n}\big\rangle \label{tran}%
\end{equation}
where $\mathcal{F}%
_{t}=\dd\mathcal{A}_{t}+\mathcal{A}_{t}\wedge \mathcal{A}_{t}$ and
$\mathcal{A}_{t}=\mathcal{\bar{A}}+t\,
\big(\mathcal{A}-\mathcal{\bar{A}}\, \big)$ is a connection which
interpolates between the two independent gauge potentials $\mathcal{A}$ and
$\mathcal{\bar{A}} $. It is easy to check that the Chern--Simons form can be
recovered in the special limit $\mathcal{\bar{A}}=0$; in contrast,
for dynamical gauge potentials the transgression form $Q_{\mathcal{A}
\leftarrow\mathcal{\bar{A}}}^{\left(  2n+1\right)  }$ is a globally
defined differential form on $\mathcal{M}$. Transgression forms satisfy two
important properties:
\begin{itemize}
\item {Triangle equation}:%
\begin{equation}
Q_{\mathcal{A}\leftarrow\mathcal{\bar{A}}}^{\left(  2n+1\right)
}=Q_{\mathcal{A}\leftarrow\mathcal{\tilde{A}}}^{\left(  2n+1\right)
}-Q_{\mathcal{\bar{A}}\leftarrow\mathcal{\tilde{A}}}^{\left(  2n+1\right)
}-\dd Q_{\mathcal{A}\leftarrow\mathcal{\bar{A}}\leftarrow\mathcal{\tilde{A}}%
}^{\left(  2n\right)  } \ . \label{triangleq}
\end{equation}
\item {Antisymmetry}:
\begin{equation}
Q_{\mathcal{A}\leftarrow\mathcal{\bar{A}}}^{\left(  2n+1\right)  }%
=-Q_{\mathcal{\bar{A}}\leftarrow\mathcal{A}}^{\left(  2n+1\right)  } \ .
\end{equation}
\end{itemize}
The first property splits a transgression form into the sum of two transgression
forms depending on an intermediate connection $\mathcal{\tilde{A}}$\ plus an exact
form with \cite{Iza05}
\begin{equation}
Q_{\mathcal{A}\leftarrow\mathcal{\bar{A}}\leftarrow\mathcal{\tilde{A}}%
}^{\left(  2n\right)  }:= n\, \left(  n+1\right) \, \int_{0}^{1}\, \dd
t \ \int_{0}%
^{t}\, \dd s \ \big\langle \big(  \mathcal{A}-\mathcal{\bar{A}}\,\big)  \wedge\big(
\mathcal{\bar{A}}-\mathcal{\tilde{A}}\, \big)  \wedge\mathcal{F}_{st}%
^{n-1}\big\rangle \label{traingle}%
\end{equation}
where $\mathcal{F}_{st}=\dd\mathcal{A}_{st}+\mathcal{A}_{st}%
\wedge\mathcal{A}_{st}$ with $\mathcal{A}_{st}=\mathcal{\tilde{A}}+s\,\big(
\mathcal{A}-\mathcal{\bar{A}}\, \big)  +t\, \big(  \mathcal{\bar{A}%
}-\mathcal{\tilde{A}}\, \big)  $. Without any loss of generality, it is always
possible to impose $\mathcal{\tilde{A}}=0$ so that the transgression
form becomes the
difference of two Chern--Simons forms plus a boundary term%
\begin{equation}
Q_{\mathcal{A}\leftarrow\mathcal{\bar{A}}}^{\left(  2n+1\right)  }%
=\mathrm{L}_{\mathrm{CS}}^{\left(  2n+1\right)  }\left(  \mathcal{A}\right)
-\mathrm{L}_{\mathrm{CS}}^{\left(  2n+1\right)  }\left(  \mathcal{\bar{A}%
}\, \right)  -\dd\, \mathrm{B}^{\left(  2n\right)  }\left(  \mathcal{A},\mathcal{\bar
{A}}\, \right) \label{trans1}%
\end{equation}
where $\mathrm{B}^{\left(  2n\right)  }\left(  \mathcal{A},\mathcal{\bar{A}
}\, \right) =Q_{\mathcal{A}\leftarrow\mathcal{\bar{A}}\leftarrow
0}^{(2n)} $. This equation illustrates that the global form
$Q_{\mathcal{A}\leftarrow\mathcal{\bar{A}}}^{\left(  2n+1\right)  }$
can be interpreted as a globalization of the sheaf of local
Lagrangians $\mathrm{L}_{\mathrm{CS}}^{\left(  2n+1\right)  }\left(
  \mathcal{A}\right)=Q_{\mathcal{A}\leftarrow 0}^{\left(  2n+1\right)
}$, with $\bar{\mathcal{A}}$ regarded as a background reference
connection; in particular, on a local trivialization of $\mathcal{P}$
one can set $\bar{\mathcal{A}}=0$.

\subsection{Topological gravity actions}

Let now $\mathcal{M}$ be a manifold of dimension $d=2n+1$ with
boundary $\partial\mathcal{M}$. Let $\mathcal{A}$ and $\mathcal{\bar{A}}$ be the linear and nonlinear one-form gauge potentials both taking values in
the Lie algebra $\mathfrak{g}= \mathfrak{iso}(2n,1)$. In the following we assume
that both gauge potentials can be obtained as the pull-back by a local
section $\sigma$ of a one-form connection $\theta$ defined on a
nontrivial principal $G$-bundle $\mathcal{P}$ over $\mathcal{M}$. This
means that they are related by a gauge transformation which we take to
be given by $g=\e^{-\phi^a \,\mathsf{P}_a} \in G/H$, where
$H=SO(2n,1)$ is the Lorentz subgroup. 

From eq.~$\left(  \ref{trans1}\right)  $ we see that the transgression action
for a manifold $\mathcal{M}$ with boundary $\partial\mathcal{M}$ is given by
\begin{equation}
\mathrm{S}_{\mathrm{T}}^{(2n+1)}[  \mathcal{A}%
,\mathcal{\bar{A}}\, ]  =\kappa\, \int_{\mathcal{M}}\,
\mathrm{L}_{\mathrm{CS}%
}^{(2n+1)  }(\mathcal{A})  -\kappa\, \int_{\mathcal{M}
}\, \mathrm{L}_{\mathrm{CS}}^{(2n+1)}(\mathcal{\bar{A}\,
})  -\kappa\, \int_{\partial\mathcal{M}}\, \mathrm{B}^{(2n)
}(\mathcal{A},\mathcal{\bar{A}}\, ) \ . \label{tact}%
\end{equation}
If the $G$-bundle $\mathcal{P}$ is nontrivial, then eq.~$(\ref{tact})$ can
be written more precisely by covering $\mathcal{M}$ with local charts. This explains the introduction of the second gauge potential $\mathcal{\bar{A}}$ such that in the overlap of two charts
the connections are related by a gauge transformation of the form in
eq.~$(\ref{gtc})$ and the overlap contributions cancel; in this setting the coset element $g \in G/H$ is
interpreted as a transition function determining the nontriviality of
$\mathcal{P}$ \cite{Anabalon:2006fj}.

Now we construct transgression actions for the Poincar\'e group using
the Lagrangian of eq.~$(\ref{cslin})$ and its nonlinear representation
in eq.~$(\ref{csnonlin})$. 
In this case the boundary term $\mathrm{B}^{(2n)  }(  \mathcal{A}%
,\mathcal{\bar{A}}\, ) $ defined by eq.~$\left(  \ref{traingle}%
\right)  $ reads%
\begin{equation}
\mathrm{B}^{(2n)  }(  \mathcal{A},\mathcal{\bar{A}}\, )
= n\, \int_{0}^{1}\, \dd t \ t^{n}\, \epsilon_{a_{1} \cdots
  a_{2n+1}}\, R_{t}^{a_{1}a_{2}}
\wedge \cdots \wedge R_{t}^{a_{2n-3}a_{2n-2}}\wedge\omega^{a_{2n-1}a_{2n}}\wedge
D_{\omega}\phi^{a_{2n+1}} \ . \label{borde}%
\end{equation}
Inserting eqs.~$\left(  \ref{cslin},\ref{csnonlin}\right)  $ and eq.~$\left(
\ref{borde}\right)  $ into eq.~$\left(  \ref{tact}\right)  $ we get%
\begin{equation}
\mathrm{S}_{\mathrm{T}}^{\left(  2n+1\right)  }\left[  \mathcal{A}%
,\mathcal{\bar{A}}\, \right]    =\kappa\, \int_{\mathcal{M}}\, \epsilon
_{a_{1} \cdots a_{2n+1}}\, R^{a_{1}a_{2}}\wedge \cdots \wedge R^{a_{2n-1}a_{2n}}\wedge
D_{\omega}\phi^{a_{2n+1}} \ ,
\end{equation}
which is a boundary term because of the Bianchi identity
$D_\omega R^{ab}=0$ and Stokes' theorem. This motivates the writing
\begin{equation}
\mathrm{S}^{\left(  2n\right)  }\left[  \omega,\phi\right]  =\kappa\, \int
_{\partial\mathcal{M}}\, \epsilon_{a_{1} \cdots a_{2n+1}}\, R^{a_{1}a_{2}}\wedge \cdots \wedge R^{a_{2n-1}%
a_{2n}}\, \phi^{a_{2n+1}} \label{topg}
\end{equation}
as an action
principle in one less dimension which corresponds to
$2n$-dimensional topological Poincar\'e gravity. Our derivation
can be regarded as a holographic principle in the
sense that the transgression action in eq.~$(\ref{tact})$ collapses to
its boundary contribution once we consider gauge connections taking
values in the Lie algebras associated to the linear and nonlinear
realizations of the Poincar\'e group. The topological action of
eq.~$(\ref{topg})$ is the action of a gauged WZW
model~\cite{Ana07}; this is because the transformation law for
the nonlinear gauge fields has the same form as a gauge transformation
from eq.~$(\ref{gtc})$ with gauge element $g=\e^{-\phi^{a}\,
  \mathsf{P}_a} \in \frac{ISO(2n,1)}{SO(2n,1)}$
\cite{Salgado:2013pva}. 

Recall that the nonlinear realization prescribes a transformation law
for the field $\phi$ under local translations given by $\delta
\phi^a=\rho^a$. This transformation breaks the symmetry of
eq.~$(\ref{topg})$ from $ISO(2n,1) $ to $ SO(2n,1)$; this is due to
the fact that the transformation law of the coset field $\phi$ under
local translations is not a proper adjoint transformation (see
eq.~$(\ref{nonl3})$). 
 
The variation of the action in eq.~$(\ref{topg})$ leads to the field
equations
\begin{align}
\epsilon_{abc a_{1} \cdots a_{2n-2}}\,D_\omega\phi^c \wedge R^{a_{1}a_{2}} \wedge \cdots  \wedge R^{a_{2n-3}%
a_{2n-2}} & =0 \ , \label{eomtop1} \\[4pt]
\epsilon_{c a_{1} \cdots a_{2n}}\, R^{a_{1}a_{2}} \wedge \cdots \wedge
R^{a_{2n-1}a_{2n}}  & =0 \ . \label{eomtop2}
\end{align}

Note that one can always use a gauge transformation to rotate to a
frame in which $\phi^1= \cdots = \phi^{a_{2n}}=0$ and $\phi^{a_{2n+1}}
:= \phi$. This choice breaks the gauge symmetry to the residual gauge
symmetry preserving the frame, which is a subgroup $SO(2n-1,1)
\hookrightarrow SO(2n,1)$; this is just the usual Lorentz symmetry
in $2n$ dimensions. If in addition one imposes the condition
$\omega^{a,{2n+1}}=0$ for $a=1, \ldots ,2n$, then gauge invariance of
eq.~(\ref{topg}) is also preserved. 

\newsection{Topological supergravity} \label{sect5}

\subsection{Three-dimensional supergravity}

Supergravity in three dimensions \cite{Des83,Ach86} can be formulated
as a Chern--Simons theory for the Poincar\'e supergroup
\cite{Banh96a}; our spinor conventions are summarized in Appendix~\ref{app2}. In the case of $\mathcal{N}=1$ supersymmetry, the model is
described by the action%
\begin{equation}
\mathrm{S}^{\left(  3\right)  }\left(  \mathcal{A}\right)
=\kappa\, \int_{\mathcal{M}}\, \mathrm{L}_{\mathrm{CS}}^{\left(  3\right)  }\left(
\mathcal{A}\right) = \kappa\, \int_{\mathcal{M}}\, \big(
\epsilon_{abc}\, R^{ab}\wedge
e^{c}-\ii\bar{\psi}\wedge D_\omega\psi\big) -\frac{\kappa}{2}\, \int_{\partial\mathcal{M}}\, \epsilon_{abc}\, \omega^{ab}\wedge
e^{c}\label{sulinact}%
\end{equation}
where ${\psi}$ is a two component Majorana spinor
one-form. This action
is invariant (up to boundary terms) under Lorentz rotations, Poincar\'e
translations and $\mathcal{N}=1$ supersymmetry transformations. The gauge fields $e^a$, $\omega^{ab}
$ and $\bar{\psi}$ transform as components of a gauge connection
valued in the $\mathcal{N}=1$ supersymmetric extension of Poincar\'e
algebra in three dimensions given by
\begin{equation}
\mathcal{A}=\ii e^{a}\,  \mathsf{P}_{a}+\mbox{$\frac{\ii}{2}$}\,
\omega^{ab}\, \mathsf{J}_{ab}%
+\bar{\psi}\, \mathsf{Q} \ .
\end{equation}
This algebra contains, in addition to the bosonic commutation relations, the
supersymmetric relations
\begin{equation}
[  \mathsf{J}_{ab},\mathsf{Q}_{\alpha}] =-
\mbox{$\frac{\ii}{2}$}\,  \left(\Gamma_{ab}
\right)_{\alpha}^{~\beta}\, \mathsf{Q}_{\beta} \qquad \mbox{and}
\qquad \left\{  \mathsf{Q}_{\alpha},\mathsf{Q}_{\beta}\right\} =\left(
\Gamma^{a}\right)  _{\alpha \beta}\, \mathsf{P}_{a} \ .
\end{equation}
Here $\Gamma_{ab}=\left[  \Gamma_{a},\Gamma_{b}\right]  $, and the set of
gamma-matrices $\Gamma_{a}$ with $a=1,2,3$ defines a representation of the
Clifford algebra in $2+1$ dimensions; then
$D_\omega\psi:=\dd\psi+\frac14\, \omega^{ab}\wedge \Gamma_{ab}\psi$
is the Lorentz covariant derivative in the spinor representation. Under an infinitesimal
gauge transformation with parameter $\lambda=\ii \rho^{a}\, \mathsf{P}_{a}+\frac
{\ii}{2}\, \kappa^{ab}\, \mathsf{J}_{ab}+\bar{\varepsilon}\, \mathsf{Q}$, the gauge fields transform
as
\begin{align}
\delta e^{a} &  =-D_{\omega}\rho^{a}+\kappa_{~b}^{a}\,
e^{b} \ ,
\nonumber \\[4pt]
\delta\omega^{ab} &  =-D_{\omega}\kappa^{ab} \ ,  \nonumber \\[4pt]
\delta\bar{\psi} &  =-D_\omega\bar{\varepsilon}-\mbox{$\frac{1}{4}$}\,
\kappa^{ab}\, \bar{\psi
}\, \Gamma_{ab} \ . 
\end{align}
These transformations leave the action of eq.~$\left(  \ref{sulinact}\right)  $ invariant
modulo boundary terms.

\subsection{Supersymmetric SWGN formalism}

The supersymmetric Stelle--West--Grignani--Nardelli formalism is treated in ref.~\cite{Sal03a} where
the nonlinear realization of the supersymmetric \textrm{AdS} group in
three dimensions is considered. Here we consider the
nonlinear realization of the three-dimensional $\mathcal{N}=1$ Poincar\'e supergroup \cite{Sal01}.

Let $G$ denote the Poincar\'e supergroup generated by $\left\{  \mathsf{J}%
_{ab},\mathsf{P}_{a},\mathsf{Q} \right\}  $. It is convenient
to decompose $G$ into two subgroups: The Lorentz subgroup $L=SO(2,1)$ generated by
$\left\{  \mathsf{J}_{ab}\right\}  $ as the stability subgroup, and the
Poincar\'e subgroup $H=ISO(2,1)$ generated by $\left\{  \mathsf{J}_{ab},\mathsf{P}%
_{a}\right\}  $. We introduce a coset field associated to each generator in
the coset space $G/L$ through
$\bar{\chi}\,\mathsf{Q}$ and $\phi^{a}\, \mathsf{P}_{a}$. 
Let us write eq.~$\left(  \ref{nonl2}\right)  $ in the form%
\begin{equation}
g_{0}\, \e^{-\bar{\chi}\,\mathsf{Q}}\,  \e^{-\phi\text{\textperiodcentered}\mathsf{P}%
}= \e^{-\bar{\chi}^{\prime}\,\mathsf{Q}}\, 
\e^{-\phi^{\prime}\text{\textperiodcentered}\mathsf{P}}\, l_{1}
\end{equation}
with $l_1\in L$.
Multiplying on the right by $\e^{\phi\text{\textperiodcentered}\mathsf{P}}$ we
get%
\begin{equation}
g_{0}\, \e^{-\bar{\chi}\,\mathsf{Q}}\,  =\e^{-\bar{\chi}^{\prime}%
\,\mathsf{Q}}\,  h_{1}\qquad \mbox{and}
\qquad 
h_{1}\, \e^{-\phi\text{\textperiodcentered}\mathsf{P}} =\e^{-\phi^{\prime
}\text{\textperiodcentered}\mathsf{P}}\, l_{1}%
\end{equation}
with $h_{1}=\e^{-\phi^{\prime}\text{\textperiodcentered}\mathsf{P}}\,
l_{1}\, \e^{\phi
\text{\textperiodcentered}\mathsf{P}}\in H$. To obtain the transformation law of
the coset fields, we write these expressions in infinitesimal form%
\begin{align}
\e^{\bar{\chi}\,\mathsf{Q}} \left(  g_{0}-1\right)  \e^{-\bar{\chi}\, \mathsf{Q}}-\e^{\bar{\chi}\,\mathsf{Q}}\, \delta\big(
\e^{-\bar{\chi}\, \mathsf{Q}}\big) &  =h_{1}-1 \ , \label{suinf1}\\[4pt]
\e^{\phi\text{\textperiodcentered}\mathsf{P}}\, \left(  h_{1}-1\right)\,
\e^{-\phi\text{\textperiodcentered}\mathsf{P}}-\e^{\phi\text{\textperiodcentered
}\mathsf{P}}\, \delta \e^{-\phi\text{\textperiodcentered}\mathsf{P}} &
=l_{1}-1 \ , \label{suinf2}%
\end{align}
where $h_{1}=h_{1}\left(  \bar{\chi},\bar{\varepsilon}
,\rho,\kappa\right)  $ and $l_{1}=l_{1}\left(  \bar{\chi},\phi
,\bar{\varepsilon},\rho,\kappa\right)  $. Inserting $g_{0}%
-1=-\ii\rho^{a}\, \mathsf{P}_{a}-\frac{\ii}{2}\,
\kappa^{ab}\,\mathsf{J}_{ab}-\bar{\varepsilon}\, \mathsf{Q} $,
$h_{1}-1=-\ii \rho^{a}\, \mathsf{P}_{a}-\frac{\ii}{2}\kappa^{ab}\,\mathsf{J}_{ab}$ and $l_{1}-1=-\frac{\ii}%
{2}\, \kappa^{ab}\, \mathsf{J}_{ab}$ into eqs.~$\left(  \ref{suinf1}\text{,}%
\ref{suinf2}\right)  $, we find the symmetry transformations for the
coset fields%
\begin{align}
\delta\phi^{a} &  =\rho^{a}+\mbox{$\frac{\ii}{2}$}\,
\bar{\varepsilon}\, \Gamma^{a}\chi -\kappa_{~c}^a\, \phi^c \ , \label{nltf1}\\[4pt]
\delta\bar{\chi} &  = \mbox{$\frac{1}{4}$}\, \bar{\chi}\,
\kappa^{ab}\, \Gamma_{ab}+\bar{\varepsilon} \ . \label{nltf3}%
\end{align}
The relations between the linear and nonlinear gauge fields can be obtained
from eq.~$\left(  \ref{gtc}\right)  $. With $g=\e^{-\bar{\chi}%
\,\mathsf{Q}}\, \e^{-\phi\text{\textperiodcentered}\mathsf{P}}$
we get%
\begin{align}
V^{a} &  =e^{a}-D_\omega\phi^{a}-\mbox{$\frac{\ii}{2}$}\, D_\omega\bar{\chi} \,\Gamma^{a} \chi+\ii\bar
{\chi}\,   \Gamma^{a} \psi
 \ , \label{nlf1}\\[4pt]
W^{ab} &  =\omega^{ab} \ , \label{nlf2}\\[4pt]
\bar{\Psi} &
=\bar{\psi}-D_\omega\bar{\chi} \ . \label{nlf3} 
\end{align}
In this way the action for supergravity in three dimensions
written in terms of nonlinear fields reads%
\begin{equation}
\mathrm{S}^{\left(  3\right)  }\left(  \mathcal{\bar{A}%
}\, \right) =\kappa\, \int_{\mathcal{M}}\, \mathrm{L}_{\mathrm{CS}}^{\left(  3\right)
}\left(  \mathcal{\bar{A}}\, \right)  = \kappa\,
\int_{\mathcal{M}}\, \big( \epsilon
_{abc}\, R^{ab}\wedge V^{c}-\ii 
\bar{\Psi}\wedge D_\omega\Psi \big) -\frac{\kappa}{2}\, \int_{\partial\mathcal{M}}\, \epsilon_{abc}\,
\omega^{ab} \wedge
V^{c} \ . \label{sunolinact}%
\end{equation}

\subsection{Topological supergravity in two dimensions}

In complete analogy with the bosonic case, we now construct a transgression
action for the Poincar\'e supergroup in three dimensions.
Inserting eq.~$\left(  \ref{sulinact}\right)  $ and eq.~$\left(
\ref{sunolinact}\right)  $ in eq.~$\left(  \ref{tact}\right)  $ with
\begin{equation}
\mathrm{B}^{\left(  2\right)  }\left(
  \mathcal{A},\mathcal{\bar{A}}\, \right)
= \mbox{$-\frac{1}{2}$}\, \epsilon_{abc}\, \omega^{ab}\wedge \big(  D_\omega\phi^{c}+\ii D_\omega\bar{\chi
}\, \Gamma^{c} \chi-\ii\bar{\chi}\,  \Gamma^{c}
\psi \big)  
 +\ii\bar{\psi}\wedge D_\omega\chi
\end{equation}
we obtain%
\begin{equation}
\mathrm{S}^{\left(  2\right)  }\left[\omega,\phi;\bar{\psi},\chi\right]
=\kappa\, \int_{\partial\mathcal{M}}\, \big( \epsilon_{abc}\, R^{ab}\,
 \phi^{c} -2\ii\bar{\psi}\wedge D_\omega\chi
 \big) \ . \label{supact}%
\end{equation}
This action corresponds to the supersymmetric extension of topological
gravity in two dimensions proposed by
ref.~\cite{Chamseddine:1989yz}. As in the
purely bosonic case, supersymmetry is broken to the Lorentz symmetry
$SO(2,1)$ because of the nonlinear transformation laws in
eqs.~(\ref{nltf1},\ref{nltf3}); however, the action is invariant
under the full supersymmetry if one prescribes the correct
transformation laws for the coset fields $\bar{\chi}, \phi$
instead of considering the symmetries dictated by the nonlinear
realization. The variation of the action in eq.~$(\ref{supact})$ leads
to the field equations
\begin{equation}
 \epsilon_{abc} \, \left(
   D_\omega\phi^{c}-\ii\bar{\psi}\,\Gamma^{c}\chi \right) =0 \ ,
 \qquad D_{\omega} \chi =0 = D_{\omega}{\bar{\psi}} \qquad \mbox{and}
 \qquad \epsilon_{abc}\, R^{ab}=0 \ .
\end{equation}

\newsection{WZW model for the gauged Maxwell algebra}  \label{sect6}

\subsection{Maxwell algebra and Chern--Simons gravity}

The Maxwell algebra is a noncentral extension of the Poincar\'e algebra by a rank two tensor $\mathsf{Z}_{ab}=-\mathsf{Z}_{ba}$ such that
 \begin{equation}                                 
 \left[ \mathsf{P}_{a},\mathsf{P}_{b}\right]=\mathsf{Z}_{ab} \qquad
 \mbox{and} \qquad
\left[ \mathsf{J}_{ab},\mathsf{Z}_{cd}\right] =\eta _{bc}\, \mathsf{Z}%
_{ad}+\eta _{ad}\, \mathsf{Z}_{bc}-\eta _{ac}\, \mathsf{Z}_{bd}-\eta
_{bd}\, \mathsf{Z%
}_{ac} \ .
\end{equation}
It describes the symmetries of a particle moving in an
electromagnetic background  \cite{Bacry:1970ye, Schrader:1972zd}. It is argued in
ref.~\cite{deAzcarraga:2010sw} that gauging the Maxwell algebra leads to new contributions to the
cosmological term in Einstein gravity. In this section we explore the
implications of the gauged Maxwell algebra in the context of Chern--Simons
gravity. In particular, we consider the three-dimensional case because
it is in
this dimension where Einstein gravity and Chern--Simons gravity are
classically equivalent. This motivates the construction of the
corresponding gauged
WZW model in two dimensions.

In order to construct the Chern--Simons gravitational Lagrangians, on
the one hand we need to \textit{gauge} the Maxwell algebra, while on the other hand we
need to specify the non-zero components of the invariant tensor. Gauging the
Maxwell algebra is straightforward. Consider a connection one-form
$\mathcal{A}$ taking values in the Maxwell algebra, which can be
expanded as%
\begin{equation}
\mathcal{A}=e^{a}\, \mathsf{P}_{a}+\mbox{$\frac{1}{2}$}\, \omega
^{ab}\, \mathsf{J}_{ab}+\mbox{$\frac{%
1}{2}$}\, \sigma ^{ab}\, \mathsf{Z}_{ab}
\end{equation}
where $e^{a}$ and $\omega ^{ab}$ are the standard vielbein and spin
connection gauge fields, and we introduce an additional rank two
antisymmetric one-form $\sigma^{ab}=-\sigma^{ba}$ as the gauge field
corresponding to the generator $\mathsf{Z}_{ab}$. The associated invariant tensors are a
little bit more involved; they can be
obtained as an $S$-expansion starting from the \textrm{AdS}
algebra in three dimensions. $S$-expansions consist of systematic Lie
algebra enhancements which enlarge symmetries. They have the nice property
that they provide the right invariant tensor of the expanded algebra
\cite{Iza06b}, which is a key ingredient in the evaluation of
Chern--Simons forms; in Appendix~\ref{app3} we show that the Maxwell algebra can be obtained as an $S$-expansion of the AdS algebra. In
three dimensions the resulting invariant tensors for the Maxwell algebra
are found to be               
\begin{align}
\left\langle \mathsf{J}_{ab}\, \mathsf{J}_{cd}\right\rangle  &=\alpha
_{0}\, \left( \eta _{ad}\, \eta _{bc}-\eta _{ac}\, \eta _{bd}\right)
\ ,  \label{tens1} \\[4pt]
\left\langle \mathsf{J}_{ab}\, \mathsf{P}_{c}\right\rangle  &=\alpha
_{1}\, \epsilon _{abc} \ , \label{tens2} \\[4pt]
\left\langle \mathsf{J}_{ab}\, \mathsf{Z}_{cd}\right\rangle  &=\alpha
_{2}\, \left( \eta _{ad}\, \eta _{bc}-\eta _{ac}\, \eta _{bd}\right)
\ , \label{tens3}  \\[4pt]
\left\langle \mathsf{P}_{a}\, \mathsf{P}_{b}\right\rangle  &=\alpha
_{2}\, \eta_{ab} \ , \label{tens4}
\end{align}%
where $\alpha _{i}$, $i=0,1,2$ are arbitrary constants. 

With this data, one can show that the Chern--Simons gravity action for the
Maxwell algebra is given by%
\begin{align}    
\mathrm{S}_{\mathrm{CS}}^{(3)}(\mathcal{A})=&\ \kappa \,
\int_{\mathcal{M}} \, \bigg(
\frac{\alpha_{0}}{2}\, \omega_{~b}^a \wedge \Big( \dd\omega _{~c}^{b}+\frac{2%
}{3}\, \omega _{~d}^{b}\wedge \omega _{~c}^{d}\Big) +\alpha _{1}\, \epsilon
_{abc}\, R^{ab}\wedge e^{c}  \label{maxcs}\\
& \qquad\qquad +\alpha _{2}\, \left( T^{a}\wedge e_{a}+R_{~c}^{a}\wedge \sigma
_{~a}^{c}\right) -\dd\Big( \, \frac{\alpha _{1}}{2}\, \epsilon _{abc}\, \omega
^{ab}\wedge e^{c}+\frac{\alpha _{2}}{2}\, \omega _{~b}^{a}\wedge \sigma
_{~a}^{b}\, \Big) \bigg) \ , \nonumber
\end{align}%
where $T^a=D_\omega e^a$ is the torsion two-form. The resulting theory
contains three sectors governed by the different values of the
coupling constants $\alpha_i$. The first term is the gravitational
Chern--Simons Lagrangian~\cite{Wit88} while the second term is the
usual Einstein--Hilbert Lagrangian. The sector proportional to
$\alpha_2$ contains the torsional term plus a new coupling between the
gauge field $\sigma^{ab}$ and the Lorentz curvature. Up to boundary
terms, the action of eq.~$(\ref{maxcs})$ is invariant under the local
gauge transformations%
\begin{align}
\delta e^{a} &=-D_{\omega }\rho ^{a}+\kappa _{~b}^{a}\, e^{b} \ ,
\nonumber \\[4pt]
\delta \omega ^{ab} &=-D_{\omega }\kappa ^{ab} \ , \nonumber \\[4pt]
\delta \sigma ^{ab} &= -D_{\omega }\tau ^{ab}-2e^{a}\, \rho ^{b}-2\omega
_{~c}^{a} \tau ^{cb}+2\kappa _{~c}^{a}\, \sigma ^{cb} \ .
\end{align}

The variation of eq.(\ref{maxcs}) leads to the following equations of motion
\begin{align}
\alpha_{0}R_{ab}+\alpha
_{1}\epsilon_{abc}T^{c}-\alpha_{2}\left(  e_{a} \wedge e_{b}+\frac{1}{2}D_{\omega}\sigma_{ab}\right)  =0 ,  \label{eom1}\\
\alpha_{1}\epsilon_{abc}%
R^{ab}+2\alpha_{2}T_{c}=0 \label{eom2},\\
\alpha_{2}R_{ab}=0 \label{eom3}.
\end{align}
Substituting eq.$(\ref{eom3})$, with $\alpha_2 \neq 0$, into eq.$(\ref{eom2})$ we get $T^a=0$. Substituting again into eq.$(\ref{eom1})$ we finally get
\begin{align}
R_{ab} &  =0,\label{flatr}\\
T_{c} &  =0,\label{flatt}\\
D_{\omega}\sigma_{ab} +2e_{a} \wedge e_{b}&  =0. \label{flats}%
\end{align}
Thus, according to eq.$(\ref{flatr}, \ref{flatt})$, the three dimensional Chern--Simons action for the Maxwell algebra describes a flat geometry. The new feature of the theory comes from eq.$(\ref{flats})$ which can be interpreted as the coupling of a matter field $\sigma$ to the flat three dimensional space.
\subsection{WZW model}                      
          
The Maxwell group $G$ contains the Lorentz subgroup $H$ generated by $\{%
\mathsf{J}_{ab}\}$ and the coset $G/H$ generated by $\left\{ \mathsf{P}_{a},%
\mathsf{Z}_{ab}\right\} $. Under gauge transformations, the gauge field
transforms according to eq.~$(\ref{gtc})$.                                                     
Let us now perform a gauge transformation with gauge element $g \in G/H $ given by
\begin{equation}
g=\e^{-\frac{1}{2}\, h^{ab}\, \mathsf{Z}_{ab}}\, \e^{-\phi ^{a}\,
  \mathsf{P}_{a}} \ .
\end{equation}           
In terms of gauge fields, eq.~$(\ref{gtc})$ reads
\begin{equation}      
V^{a}\, \mathsf{P}_{a}+\mbox{$\frac{1}{2}$}\, W^{ab}\,
\mathsf{J}_{ab}+ \mbox{$\frac{1}{2}$}\, \Sigma^{ab} \,
\mathsf{Z}_{ab}=\e^{\phi ^{a}\, \mathsf{P}_{a}}\, \e^{\frac{1}{2}\,
  h^{ab}\, \mathsf{Z}%
_{ab}}\, \left(\dd+\mathcal{A}\right) \e^{-\frac{1}{2}\, h^{ab}\,
\mathsf{Z}_{ab}}\, \e^{-\phi ^{a}\,
\mathsf{P}_{a}}
\end{equation}                           
and it is straightforward to show using the commutation relations that  
\begin{align}                 
V^{a} &=e^a -D_{\omega}\phi^a \ , \nonumber \\[4pt]
W^{ab} &=\omega ^{ab} \ , \nonumber \\[4pt]
\Sigma ^{ab} &=2\phi ^{a}\, e^{b}+\sigma ^{ab}-\phi ^{a}\,
D_\omega\phi ^{b}-D_\omega h^{ab} \ .
\end{align}%
The final step is to compute the transgression form from
eq.~$(\ref{trans1})$ with the result
\begin{align}
\mathrm{B}^{(2)}(\mathcal{A},\mathcal{\bar{A}}\, )=& \ \alpha
_{2}\,e^{a}\wedge \left( e_{a}-D_\omega \phi _{a}\right) -\mbox{$\frac{%
\alpha _{1}}{2}$}\, \epsilon _{abc}\, \omega ^{ab}\wedge D_\omega\phi
^{c} \nonumber \\ & + \mbox{$\frac{\alpha _{2}}{2}$}\, \omega
_{~c}^{a}\wedge \left( 2\phi ^{c}\, e_{a}-\phi ^{c}\, D_\omega\phi
_{a}-D_\omega h_{~a}^{c}\right) \ .
\end{align}
The resulting action is a boundary term which corresponds to the gauged
WZW action associated to the Maxwell algebra. As previously, we propose it as a Lagrangian in one less dimension%
\begin{equation}
\mathrm{S}^{(2)}\left[\omega, \phi, h, e \right]=\kappa\,
\int_{\partial \mathcal{M}}\, \big(\alpha _{1}\, \epsilon _{abc}\, R^{ab}\, \phi ^{c}+\alpha
_{2}\, \left( R_{~c}^{a}\wedge h_{~a}^{c}-e^{a}\wedge e_{a}\right)
\big) \ . \label{wzwmax}
\end{equation}
This action generalizes the topological action for gravity from eq.~$(\ref{topg})$. However, it is interesting to precise that both actions are classically equivalent, on-shell.  

The variation of eq.$(\ref{wzwmax})$ gives the following equations of motion
\begin{align}
\alpha_{2}D_{\omega}h_{ab}-\alpha_{1}\epsilon_{abc}D_{\omega}\phi^{c}   =0 ,& \label{wzwmax1}\\
e_{a}   =0 =R_{ab} .& \label{wzwmax2}
\end{align}
Now, making the following redefinition $\bar{\phi}^c=\alpha_2 \epsilon^{cjk}h_{jk}-2\alpha_1 \phi^c$, it is direct to show that eq.$(\ref{wzwmax1})$ satisfy $\epsilon_{abc}D_{\omega}\bar{\phi}^c=0$, which corresponds, together with eq.$(\ref{wzwmax2})$, to the field equations for the topological gravity theory in the $n=1$ case eq.$(\ref{eomtop1}, \ref{eomtop2})$. Note that this equivalence is only classical. It would be interesting to investigate what are the implications of this model at the quantum level.
\subsection*{Acknowledgments}

We thank F. Izaurieta, N. Merino, A. P\'erez and E. Rodriguez for enlightening discussions and helpful comments.
The work of P.S. and O.V. was supported in part by Direcci\'{o}n de Investigaci\'{o}n,
Universidad de Concepci\'{o}n through Grant \#212.011.056-1.0 and in part by
\textrm{FONDECYT} through Grant \#$1130653$. The work of R.J.S. and
O.V. was supported in part by the Consolidated Grant \#ST/J000310/1
from the UK Science and Technology Facilities Council. The work of O.V. is
supported by grants from the Comisi\'{o}n Nacional de Investigaci\'{o}n
Cient\'{\i}fica y Tecnol\'{o}gica \textrm{CONICYT} and from the Universidad de
Concepci\'{o}n, Chile.

\appendix

\newsection{Poincar\'e-invariant Chern--Simons gravity} \label{app1}

Poincar\'e gravity in $2n+1$ dimensions can be formulated as a Chern--Simons theory 
for the gauge group 
$ISO(2n,1)$. The fundamental field is the one-form connection 
\begin{equation}
\mathcal{A}=e^a \, \mathsf{P}_a+\mbox{$\frac{1}{2}$}\, \omega^{ab}\, \mathsf{J}_{ab}
\end{equation}
with values in the Lie algebra $\mathfrak{iso}(2n,1)$ whose
commutation relations are given by
\begin{align}
\left[  \mathsf{J}_{ab},\mathsf{J}_{cd}\right] &  =\eta_{ac}\, \mathsf{J}%
_{bd}+\eta_{bd}\, \mathsf{J}_{ca}-\eta_{bc}\, \mathsf{J}_{ad}-\eta_{ad}\,
\mathsf{J}_{bc} \ , \nonumber \\[4pt]
\left[  \mathsf{J}_{ab},\mathsf{P}_{c}\right] &  =\eta_{ac}\, \mathsf{P}%
_{b}-\eta_{bc}\, \mathsf{P}_{a} \ , \nonumber \\[4pt]
\left[  \mathsf{P}_{a},\mathsf{P}_{b}\right] & =0 \ .
\end{align}
Here $\left\{  \mathsf{J}_{ab}\right\}  _{a,b=1}^{2n+1}$ generate
the Lorentz subalgebra $\mathfrak{so}(2n,1)$, $\left\{  \mathsf{P}_{a}\right\}  _{a=1}%
^{2n+1}$ generate local Poincar\'e translations and
$(\eta_{ab}) ={\rm diag}\left(  -1,1, \ldots ,1\right)  $ is a $(2n+1)$-dimensional Minkowski metric.

In order to obtain the explicit form of the action, we use the
{subspace separation method}~\cite{Iza05,Iza06a}. The subspace separation method is a
systematic procedure for computing Chern--Simons forms. This mechanism is based on the extended Cartan homotopy formula \cite{Man85} and
has the virtue that it enables one to separate the action in terms of
bulk and boundary contributions, and it splits the
Lagrangian into pieces valued on the subspace structure of the gauge algebra
which simplifies the calculations considerably. Following refs.~\cite{Iza05,Iza06a}, first we decompose the gauge algebra
into vector subspaces $\mathfrak{iso}\left(  2n,1\right)  =\mathsf{V}_{1}\oplus \mathsf{V}_{2}$ where
$\mathsf{V}_{1}={\rm Span}_{\complex}\left\{  \mathsf{J}_{ab}\right\}
$ and $\mathsf{V}_{2}={\rm Span}_{\complex}\left\{  \mathsf{P}_{a}\right\}  $. Next
we split the  gauge potential into pieces valued in each subspace of the gauge algebra
\begin{equation}
\mathcal{A}_{0}  =0 \ , \qquad 
\mathcal{A}_{1}  =\omega\qquad \mbox{and} \qquad
\mathcal{A}_{2}  =\omega+e
\end{equation}
where $\omega=\frac{1}{2}\, \omega^{ab}\, \mathsf{J}_{ab}$ and
$e=e^{a}\, \mathsf{P}_{a}$. Computing each component of the triangle
equation of eq.~$(\ref{triangleq})$ we find
\begin{align}
Q_{\mathcal{A}_{2}\leftarrow\mathcal{A}_{1}}^{\left(  2n+1\right)  }  
& =\epsilon_{a_{1}\cdots a_{2n+1}}\, R^{a_{1}a_{2}}\wedge \cdots \wedge R^{a_{2n-1}a_{2n}
}\wedge e^{a_{2n+1}} \ , \label{t1x}\\[4pt]
Q_{\mathcal{A}_{1}\leftarrow \mathcal{A}_0}^{\left(  2n+1\right)  }   & =0 \
, \label{t2x}\\[4pt] 
Q_{\mathcal{A}_{2}\leftarrow\mathcal{A}_{1}\leftarrow\mathcal{A}_0}^{\left(  2n\right)  }
& =-n\, \int_{0}^{1}\, \dd t \ t^{n}\, \epsilon_{a_{1}\cdots
  a_{2n+1}}\, R_{t}^{a_{1}a_{2}}
\wedge \cdots \wedge R_{t}^{a_{2n-3}a_{2n-2}}\wedge\omega^{a_{2n-1}a_{2n}}\wedge
e^{a_{2n+1}}  \ . \label{t3x}
\end{align}
Here we have used the fact that the only nonvanishing components of the
invariant tensor for the Poincar\'e algebra are given by
\begin{equation}
\left\langle \mathsf{J}_{a_{1}a_{2}}\cdots \mathsf{J}_{a_{2n-1}a_{2n}} \,
\mathsf{P}_{a_{2n+1}}\right\rangle =\frac{2^{n}}{n+1} \,
\epsilon_{a_{1}\cdots a_{2n+1}} \ .
\end{equation}
Inserting eqs.~(\ref{t1x}--\ref{t3x}) into
eq.~$(\ref{triangleq})$ and using $\mathrm{L}_{\mathrm{CS}}^{\left(
    2n+1\right)  }\left(  \mathcal{A}\right)
=Q_{\mathcal{A}_{2}\leftarrow\mathcal{A}_{0}}^{\left(  2n+1\right)  }$
we obtain eq.~$\left(  \ref{cslin}\right)  $.

\newsection{Spinors in three dimensions}\label{app2}

\subsection{Gamma-matrices}

In this appendix we summarise our conventions regarding the Clifford algebra and
spinors in three dimensions, following ref.~\cite{VanPro99}. The Clifford algebra is defined by%
\beq
\Gamma_{a}\, \Gamma_{b}+\Gamma_{b}\, \Gamma_{a}=2\eta_{ab}%
\eeq
where $(\eta_{ab})={\rm diag}(-1,1,1)$ and the minus sign is along the timelike direction. The
Pauli spin matrices $\sigma_{a}$ with $a=1,2,3$ provide a representation of the
Clifford algebra with signature $\left(  0,3\right)  $. Since we are
interested in fixing a representation with signature $\left(  1,2\right)  $,
all we need to do is to multiply one of the Pauli matrices by the
imaginary unit
and declare it to be $\Gamma_{1}$. An explicit representation is then given by%
\beq%
\Gamma_{1}=\ii\sigma_{2}=%
\begin{pmatrix}
0 & 1\\
-1 & 0
\end{pmatrix}
 \ , \qquad \Gamma_{2}=\sigma_{1}=%
\begin{pmatrix}
0 & 1\\
1 & 0
\end{pmatrix}
 \qquad \mbox{and} \qquad \Gamma_{3}=\sigma_{3}=%
\begin{pmatrix}
1 & 0\\
0 & -1
\end{pmatrix}
\ .
\eeq

There always exists a charge conjugation matrix $\mathcal{C}$ which in three
dimensions satisfies
\beq
\mathcal{C}^{\top}=-\mathcal{C} \qquad \mbox{and} \qquad \Gamma_{a}^{\top}=-\mathcal{C}%
\, \Gamma_{a}\, \mathcal{C}^{-1} \ .
\eeq
In the chosen basis of gamma-matrices it can be taken to be $\mathcal{C}=\sigma_{2}=-\ii\Gamma_{1}$. The charge
conjugation matrix satisfies $\mathcal{C=C}^{-1}=\mathcal{C}^{\dagger}$.

\subsection{Majorana spinors}

The minimal irreducible spinor in three dimensions is a two real component
Majorana spinor. Every Majorana spinor satisfies a reality condition which can
be established by demanding that the Majorana conjugate equals the Dirac
conjugate
\begin{equation}
\bar{\psi}:= \psi^{\top}\mathcal{C}=-\ii\psi^{\top}\Gamma_{1} \ .\label{maj1}%
\end{equation}
Spinors carry indices $\psi_{\alpha}$ and gamma-matrices act on them in such a
way that $\Gamma_{a}\psi:= \left(  \Gamma_{a}\right)  _{~\beta}^{\alpha
}\, \psi_{\alpha}$. In order to raise and lower indices, we introduce matrices
$(\mathcal{C}^{\alpha\beta})$, $(\mathcal{C}_{\alpha\beta})$ related to the
charge conjugation matrix, and we use the convention of raising and lowering
indices according to the NorthWest--SouthEast convention $\left(
\searrow\right)  $. This means that the position of the indices should appear
in that relative position as
\begin{equation}
\psi^{\alpha}=\mathcal{C}^{\alpha\beta}\, \psi_{\beta} \qquad
\mbox{and} \qquad \psi_{\alpha}%
=\psi^{\beta}\, \mathcal{C}_{\alpha\beta} \ , \label{maj2}
\end{equation}
which implies that
\beq
\mathcal{C}^{\alpha\beta}\, \mathcal{C}_{\gamma\beta}=\delta_{\gamma}^{\alpha
} \qquad \mbox{and} \qquad \text{ }\mathcal{C}_{\beta\alpha}\, \mathcal{C}^{\beta\gamma}=\delta_{\alpha
}^{\gamma} \ .
\eeq
We choose the identifications in such a way that the Majorana conjugate
$\bar{\psi}$ is written as $\psi^{\alpha}$. Comparing eq.~(\ref{maj1})
with eq.~(\ref{maj2}), one then finds
$(\mathcal{C}^{\alpha\beta})=\mathcal{C}^{\top}$ and $(\mathcal{C}_{\alpha\beta
})=\mathcal{C}^{-1}$.

\newsection{Maxwell algebra by $S$-expansion} \label{app3}

\subsection{$S$-expansions of Lie algebras}

Let $\mathfrak{g}$ be a Lie algebra and $\mathrm{S}=\left\{  \lambda_{\alpha}\right\}  $ a finite abelian
semigroup with composition law $\lambda_\alpha\cdot\lambda_\beta$. By~\cite[Theorem~3.1]{Iza06b} the direct product $\mathrm{S}\times\mathfrak{g}$ is also a Lie
algebra. There are cases in which it is possible to systematically
extract Lie subalgebras from $\mathrm{S}\times\mathfrak{g}$. For
instance, we can start by decomposing
$\mathfrak{g}$ into a direct sum of subspaces $\mathfrak{g}%
=\bigoplus_{p\in I}\, V_{p}$ where $I$ is some index set. The Lie algebra
structure of $\mathfrak{g}$ can be encoded in subsets $I_{\left(
p,q\right)  }\subset I$ according to $\left[  V_{p},V_{q}\right]
\subset\bigoplus_{r\in I_{\left(  p,q\right)  }}\, V_{r}.$ If the
semigroup $\mathrm S$ also
admits a decomposition into subsets $\mathrm{S}=\bigcup_{p\in I}\, \mathrm{S}_{p}$ satisfying $\mathrm{S}_{p}\cdot \mathrm{S}_{q}\subset
\bigcap_{r\in I_{\left(  p,q\right)  }}\, \mathrm{S}_{r},$ we say that the algebra and the semigroup decompositions are in \textit{resonance}. Then
$\mathfrak{g}_{\mathrm{R}}:=\bigoplus_{p\in I}\, \mathrm{S}_{p}\times V_{p}$ is a ``resonant
subalgebra'' of $\mathrm{S}\times\mathfrak{g}$~\cite[Theorem~4.2]{Iza06b}.

If one further has a \textit{zero} element in the
semigroup, i.e., an element $0_{\rm S}\in \mathrm{S}$ such that
$0_{\rm S}\cdot \lambda_{\alpha}=0_{\rm S}$ for all $\lambda_{\alpha
}\in \mathrm{S}$, then the whole sector
$0_{\rm S}\times\mathfrak{g}$ can be removed from the resonant subalgebra
by imposing $0_{\rm S}\times\mathfrak{g}=0$. The remaining structure, which we
refer to as the $0_{\rm S}$-reduced algebra, is still a Lie
algebra~\cite[Theorem~6.1]{Iza06b}.

\subsection{$S$-expansion of the AdS algebra}

We now show that the Maxwell algebra can be obtained by an $S$-expansion
of the AdS algebra. Let $\mathrm{S}_{\mathrm{E}}^{\left(  2\right)  }$ be the semigroup~\cite{Salgado:2014qqa}
\begin{equation}
\mathrm{S}_{\mathrm{E}}^{\left(  2\right)  }=\left\{  \lambda_{0},\lambda
_{1},\lambda_{2},\lambda_{3}\right\}
\end{equation}
with composition law%
\begin{equation}
\lambda_{\alpha}\text{\textperiodcentered}\lambda_{\beta}:= \left\{
\genfrac{.}{.}{0pt}{}{\lambda_{\alpha+\beta} \qquad \text{ if } \ \alpha+\beta
\leq3 \ ,}{\lambda_{3} \qquad \text{ if } \ \alpha+\beta>3 \ .}%
\right.
\label{SE2law}\end{equation}
Recall that the \textrm{AdS} algebra $\frg=\mathfrak{so}(d-1,2)$ in $d$ dimensions is given by%
\begin{align}
\left[\,  \bar{\mathsf{J}}_{ab},\bar{\mathsf{J}}_{cd} \, \right]    & =\eta
_{bc}\, \bar{\mathsf{J}}_{ad}+\eta_{ad}\, \bar{\mathsf{J}}_{bc}-\eta_{ac} \,
\bar{\mathsf{J}}_{bd}-\eta_{bd}\, \bar{\mathsf{J}}_{ac} \ , \label{ads1}\\[4pt]
\left[ \, \bar{\mathsf{J}}_{ab},\bar{\mathsf{P}}_{c} \, \right]    & =\eta
_{bc}\, \bar{\mathsf{P}}_{a}-\eta_{ac}\, \bar{\mathsf{P}}_{b}  \ , \label{ads2}  \\[4pt]
\left[ \, \bar{\mathsf{P}}_{a},\bar{\mathsf{P}}_{b} \, \right]    & =\bar{\mathsf{J}}_{ab} \ . \label{ads3}
\end{align}
This algebra can be decomposed into two subspaces
$
\mathfrak{g}=\mathsf{V}_{0}\oplus\mathsf{V}_{1}%
$
where $\mathsf{V}_{0}={\rm Span}_{\complex}\left\{  \bar{\mathsf{J}}_{ab}\right\}$ and $\mathsf{V}_{1}={\rm Span}_{\complex}\left\{  \bar{\mathsf{P}}_{a}\right\}$. In terms of these subspaces, the $\mathrm{AdS}$ algebra has
the structure%
\begin{equation}
\left[  \mathsf{V}_{0},\mathsf{V}_{0}\right]  \subset\mathsf{V}_{0} \
, \qquad \left[  \mathsf{V}_{0},\mathsf{V}_{1}\right]
\subset\mathsf{V}_{1} \qquad \mbox{and} \qquad \left[
  \mathsf{V}_{1},\mathsf{V}_{1}\right]  \subset\mathsf{V}_{0} \ . \label{subspa}
\end{equation}
If we now choose the partition for the semigroup
$\mathrm{S}_{\mathrm{E}}^{\left(  2\right)  }$ given by
\begin{equation}
\mathrm{S}_{0}  =\left\{  \lambda_{0},\lambda_{2}\right\}  \cup\left\{
\lambda_{3}\right\} \qquad \mbox{and} \qquad 
\mathrm{S}_{1} =\left\{  \lambda_{1}\right\}  \cup\left\{  \lambda
_{3}\right\} \ ,
\end{equation}
then this partition is resonant with respect to the structure of the
AdS algebra: Under the
semigroup multiplication law we have%
\begin{equation}
\mathrm{S}_{0}\, \text{\textperiodcentered}\, \mathrm{S}_{0}\subset\mathrm{S}_{0}
\ , \qquad \mathrm{S}_{0}\,\text{\textperiodcentered}\, \mathrm{S}_{1}\subset\mathrm{S}%
_{1} \qquad \mbox{and} \qquad \mathrm{S}_{1}\,
\text{\textperiodcentered}\, \mathrm{S}_{1}%
\subset\mathrm{S}_{0}
\end{equation}
which agrees with the decomposition in eq.~(\ref{subspa}). The resonance condition allows us to construct a resonant subalgebra
$\mathfrak{g}_{\rm R}$ defined by%
\begin{equation}
\mathfrak{g}_{\rm R}=\mathsf{W}_{0}\oplus\mathsf{W}_{1}:= \left(
\mathrm{S}_{0}\times\mathsf{V}_{0}\right)  \oplus\left(  \mathrm{S}%
_{1}\times\mathsf{V}_{1}\right) \ .
\end{equation}
Explicitly one has
\begin{align}
\mathsf{W}_{0}  & =\left\{  \lambda_{0},\lambda_{2},\lambda_{3}\right\}
\times{\rm Span}_{\complex}\left\{  \bar{\mathsf{J}}_{ab}\right\}
=:{\rm Span}_{\complex} \left\{  \mathsf{J}%
_{ab,0},\mathsf{J}_{ab,2},\mathsf{J}_{ab,3}\right\} \ , \nonumber \\[4pt]
\mathsf{W}_{1}  & =\left\{  \lambda_{1},\lambda_{3}\right\}
\times{\rm Span}_{\complex} \left\{
\bar{\mathsf{P}}_{a}\right\}  =: {\rm Span}_{\complex} \left\{  \mathsf{P}_{a,1},\mathsf{P}%
_{a,3}\right\} \ .
\end{align}
Since $\lambda_{3}$ is a zero element
in the semigroup, one can extract another subalgebra by setting $\mathsf{J}_{ab,3}=\mathsf{P}%
_{a,3}=0$; this choice still preserves the Lie algebra
structure of the residual algebra. This algebra is called a $0$-forced
resonant algebra and therefore we are left with the subspaces
\begin{align}
{\tilde{\sf W}}_{0}  ={\rm Span}_{\complex} \left\{  \mathsf{J}_{ab,0},\mathsf{J}%
_{ab,2}\right\}  \qquad \mbox{and} \qquad
{\tilde{\sf W}}_{1}  ={\rm Span}_{\complex} \left\{
  \mathsf{P}_{a,1}\right\} \ .
\end{align}
In order to obtain a presentation for the $0$-forced resonant algebra we
use eqs. (\ref{ads1}--\ref{ads3}) together with eq. (\ref{SE2law}) to compute the commutation relations and
identify
\begin{align}
\mathsf{J}_{ab}:=\mathsf{J}_{ab,0}  \ , \qquad 
\mathsf{Z}_{ab}:= \mathsf{J}_{ab,2}  \qquad \mbox{and} \qquad
\mathsf{P}_{a}:= \mathsf{P}_{a,1} 
\end{align}
to obtain the Maxwell algebra in $d$ dimensions
\begin{align}
\left[  \mathsf{J}_{ab},\mathsf{J}_{cd}\right]    & =\eta_{bc}\, \mathsf{J}%
_{ad}+\eta_{ad}\, \mathsf{J}_{bc}-\eta_{ac}\, \mathsf{J}_{bd}-\eta_{bd}\,
\mathsf{J}_{ac} \ , \nonumber \\[4pt]
\left[  \mathsf{J}_{ab},\mathsf{Z}_{cd}\right]    & =\eta_{bc}\, \mathsf{Z}%
_{ad}+\eta_{ad}\, \mathsf{Z}_{bc}-\eta_{ac}\, \mathsf{Z}_{bd}-\eta_{bd}\,
\mathsf{Z}_{ac} \ , \nonumber \\[4pt]
\left[  \mathsf{J}_{ab},\mathsf{P}_{c}\right]    & =\eta_{bc}\, \mathsf{P}%
_{a}-\eta_{ac}\, \mathsf{P}_{b} \ , \nonumber \\[4pt]
\left[  \mathsf{P}_{a},\mathsf{P}_{b}\right]    & =\mathsf{Z}_{ab} \ ,
\nonumber \\[4pt]
\left[  \mathsf{Z}_{ab},\mathsf{Z}_{cd}\right]    & =0 =
\left[  \mathsf{Z}_{ab},\mathsf{P}_{c}\right] \ .
\end{align}

\subsection{Invariant tensors}

The $S$-expansion procedure also provides the invariant tensors
associated to the expanded algebra; here we study the particular
case of $d=3$ dimensions. The invariant tensors of the AdS algebra $\mathfrak{so}(2,2)$ are
given by \cite{Hayashi:1991mu}%
\begin{align}
\left\langle \,\bar{\mathsf{J}}_{ab}\, \bar{\mathsf{J}}_{cd}\, \right\rangle  &
=\mu_{0}\, \left(  \eta_{ad}\, \eta_{bc}-\eta_{ac}\, \eta_{bd}\right)
\ , \nonumber \\[4pt]
\left\langle \, \bar{\mathsf{J}}_{ab}\, \bar{\mathsf{P}}_{c}\, \right\rangle  &
=\mu_{1}\, \epsilon_{abc} \ , \nonumber \\[4pt]
\left\langle \, \bar{\mathsf{P}}_{a}\, \bar{\mathsf{P}}_{b}\, \right\rangle  & =\mu
_{0}\, \eta_{ab} \ ,
\end{align}
where $\mu_i$, $i=0,1$ are arbitrary constants. By \cite[Theorem~7.2]{Iza06b}, the $S$-expanded tensors
are given by the formula%
\begin{equation}
\left\langle \mathsf{T}_{A,\alpha}\, \mathsf{T}_{B,\beta}\right\rangle
=\tilde{\alpha}_{\gamma}\, K_{\alpha\beta}^{~~\gamma}\, \left\langle \mathsf{T}%
_{A}\, \mathsf{T}_{B}\right\rangle \label{sinvten}
\end{equation}
where $\tilde{\alpha}_\gamma$ are also arbitrary constants, and $K_{\alpha\beta}^{~~\gamma}$ is
called a $K$-two selector which is a function with values
$1$ if $\gamma=\gamma(\alpha\, \beta)$ according to the semigroup multiplication law and
$0$ otherwise. The application of the formula in eq.~(\ref{sinvten}) for the $S$-expanded generators 
$\mathsf{J}_{ab,0}$, $\mathsf{J}_{ab,2}$ and $\mathsf{P}_{a,1}$ gives
the invariant tensors for the Maxwell algebra in
eqs.~(\ref{tens1}--\ref{tens4}), with the redefined constants
\begin{equation}
\alpha_{0}:= \tilde{\alpha}_{0}\, \mu_{0} \ , \qquad 
\alpha_{1}:= \tilde{\alpha}_{1}\, \mu_{1} \qquad \mbox{and} \qquad
\alpha_{2}:= \tilde{\alpha}_{2}\, \mu_{0} \ .
\end{equation}

\bibliographystyle{utphys}
\providecommand{\href}[2]{#2}\begingroup\raggedright\endgroup

\end{document}